\documentclass{article}

\PassOptionsToPackage{numbers, compress}{natbib}

     \usepackage[preprint]{neurips_2024}

\usepackage[utf8]{inputenc} 
\usepackage[T1]{fontenc}    
\usepackage{hyperref}       
\usepackage{url}            
\usepackage{booktabs}       
\usepackage{amsfonts}       
\usepackage{nicefrac}       
\usepackage{microtype}      
\usepackage{xcolor}         

\usepackage{amsmath, amsthm, amssymb}
\usepackage[pdftex]{graphicx}

\newcommand{\vc}[1]{\boldsymbol{#1}}

\title{Harnessing and modulating chaos to sample from neural generative models}

\author{
  Rishidev Chaudhuri\\
  Center for Neuroscience \\
  Department of Mathematics\\
  Department of Neurobiology, Physiology and Behavior\\
  University of California, Davis\\
  Davis, CA 95616\\
  \texttt{rchaudhuri@ucdavis.edu} \\
  \And
  Vivek Handebagh \\
  Center for Neuroscience \\
  Department of Physics \\
  University of California, Davis \\
  Davis, CA 95616\\
  \texttt{vshandebagh@ucdavis.edu} \\
}

\begin{document}

\maketitle

\begin{abstract}
Chaos is generic in strongly-coupled recurrent networks of model neurons, and thought to be an easily accessible dynamical regime in the brain. While neural chaos is typically seen as an impediment to robust computation, we show how such chaos might play a functional role in allowing the brain to learn and sample from generative models. We construct architectures that combine a classic model of neural chaos either with a canonical generative modeling architecture or with energy-based models of neural memory. We show that these architectures have appealing properties for sampling, including easy biologically-plausible control of sampling rates via overall gain modulation.
\end{abstract}

\section{Introduction}
Chaotic dynamics is generic in a wide variety of strongly-coupled recurrent neural network model architectures. Classic work shows that randomly-connected recurrent rate-based networks are chaotic once recurrent connection strengths cross a certain threshold \cite{sompolinsky1988}. Similar results hold for networks with structured connectivity and for spiking networks \cite{kadmon2015transition, aljadeff2015transition, harish2015asynchronous}. While it is empirically unclear when and to what extent biological neural dynamics are chaotic, neural activity shows significant population level variability that may reflect chaos, and in some settings neural dynamics can be highly sensitive to small perturbations \cite{london2010sensitivity}. The combination of theoretical arguments and experimental observations thus suggests that chaos is an easily accessible dynamical regime in the brain.

Proximity to the chaotic regime (the so-called ``edge of chaos'') has been shown to enhance computational power in both neural networks and other models of biological computation by providing long timescales and complex dynamics \cite{crutchfield1990computation, bertschinger2004real, toyoizumi2011beyond}. Chaotic dynamics themselves can offer a rich starting point for learning, providing a diverse dynamical repertoire that can be reshaped to learn complex temporal patterns \cite{sussillo2009generating, laje2013robust}. These proposed functional uses of chaos rely on moving networks out of the chaotic regime or at most keeping them slightly over the bifurcation, in a weakly chaotic regime. Could ongoing strongly chaotic dynamics play a functional role? In this study, in line with recent proposals \cite{terada2024chaotic}, we show how such chaotic dynamics might provide a natural substrate for sampling from generative models, a commonly proposed function of neural dynamics.

The ability to learn and sample from probabilistic models of the world has long been suggested as a key function of neural dynamics \cite{ackley1985learning, dayan1995helmholtz, buesing2011neural, hennequin2014fast, savin2014spatio, orban2016neural, gershman2017complex, echeveste2020cortical}. Such probabilistic models capture the distributions of events and actions in the world and are crucial to flexible intelligence. In particular, sampling from such models has been proposed as an explanation for seemingly noisy neural dynamics, and a number of architectures have been proposed that show how such sampling might occur, with a predominant focus on architectures that use stochasticity intrinsic to single neurons or synapses \cite{buesing2011neural, hennequin2014fast, savin2014spatio, orban2016neural, gershman2017complex, echeveste2020cortical}.

In this study we show how neural chaotic dynamics might be used to sample from generative models. We illustrate this idea in two ways. First, we take inspiration from common generative modeling architectures in machine learning that learn to reshape a source of variability to sample from a desired distribution \cite{Kingma2014, goodfellow2014generative, Rezende14, rezende2015variational, sohl2015deep, song2019generative, ho2020denoising}. We combine one such architecture, a Generative Adversarial Network (GAN) \cite{goodfellow2014generative}, with a canonical network model of neural chaos to show how such sampling can be implemented using chaos as the source of the variability. Second, we consider recurrent neural networks with two types of connections---one set of structured connections encodes a set of preferential states and another set of random connections promotes chaos. By modulating the balance between these two sets of connections, we show how network dynamics can sample from a distribution shaped by the structured connections and with sampling driven by chaos from the random connections. 

We moreover argue that chaos has a number of appealing features for such sampling. In particular, previous architectures for neural sampling typically rely on noise generated from single neuron fluctuations. However, typical biological neurons receive large numbers of inputs. In such large networks, in the absence of network-level amplification mechanisms noise generated from single neuron fluctuations is likely to average out, leading to slow sampling. By contrast, chaotic neural networks possess robust population-level variability. This population-level variability scales with network size, and the exponential movement of chaotic trajectories along the chaotic manifold leads to rapid mixing \cite{sompolinsky1988, clark2023dimension, engelken2023lyapunov}. Furthermore, the rate at which these networks move along the attractor and thus the rate of sampling can be modulated by a single population-level gain factor. This gain likely reflects overall excitability or neuromodulator concentration and may provide a convenient and biologically-accessible knob with which to control sampling speed.

\section{Related work}
There is a long tradition of work interpreting chaotic dynamics through a probabilistic lens, including from the perspective of sampling from some underlying invariant measure on an attracting manifold \cite{eckmann1985ergodic, gilpin2024generative}. Separately, there is also an extensive line of work seeking to understand neural variability as sampling from some probabilistic or generative model \cite{buesing2011neural, orban2016neural, echeveste2020cortical}.  In this section we focus on studies that are closely related to the present work's theme of understanding chaotic neural dynamics as sampling from generative models (and discuss our work in a more general context in the Discussion). 

A recent study by Terada \& Toyoizumi \cite{terada2024chaotic} compellingly shows how chaotic dynamics can be used to perform Bayesian inference in the context of a canonical sensory cue integration task. This work considers an initially randomly-connected recurrent neural network that is trained to output the appropriate Bayesian posterior distribution in response to two inputs representing sensory evidence cues, with one input more reliable than the other. The trained recurrent network lies in the chaotic regime, and uses these chaotic dynamics to output samples from the appropriate posterior distribution and also learns to represent the prior distribution in the absence of input. Our use of a randomly-connected recurrent neural network that generates samples via chaotic dynamics is similar to this study. Compared to this study, our work lacks a probabilistic inference task, treats learning more abstractly, and in the first (but not the second) architecture we consider separates the generation from the use of variability, which may be less biologically plausible. Distinctive features of our study include the use of significantly more complex generative models, explicitly considering how the speed of sampling could be modulated, and the proposal of hybrid architectures that combine chaotic dynamics with canonical energy-based models. 

Work by Naruse et al. \cite{naruse2019generative} uses chaotic time series data generated by lasers to provide the source of variability for a GAN, and show that the resulting GAN is able to learn and sample from a distribution of faces. This architecture is similar to the first architecture we present in this paper in Section \ref{sec:chaos_gan}, though differs in the source of chaos (we consider a recurrent neural network model instead of a laser), and in our consideration of gain modulation as a way to change sampling speed. 

The study of Jordan et al. \cite{jordan2019deterministic} also considers using a deterministic recurrent network to provide variability for sampling and thus is again similar to the first architecture we present. This study shows that a randomly-connected recurrent network of binary units can replace the assumption of neuron-intrinsic variability in a Boltzmann machine (a generative model with binary units and symmetric weights) \cite{ackley1985learning} and allows good sampling of the MNIST handwritten digits dataset. While this study does not explicitly consider chaos, it is likely that the external network they consider is in the chaotic regime.

\begin{figure}[th] 
    \begin{center}
            \includegraphics[width=\linewidth]{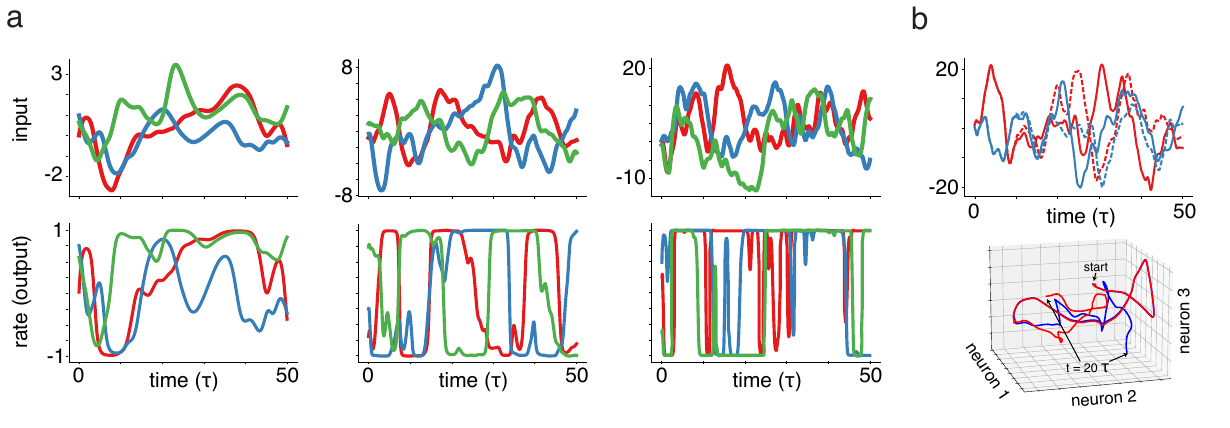}
    \end{center}
    \caption{{\bf Chaotic dynamics in a randomly-connected recurrent network of neurons.}
(a) Top row shows time-varying network input to 3 neurons (i.e., $h$ in Eq. \ref{eq:somp_chaos_net}), with each neuron shown as a different trace. Bottom row shows corresponding activity or output from these 3 neurons (i.e., $\phi(h)$). Left to right show increasing values of gain, with $g=2, 4, 10$. Trajectories become increasingly chaotic as gain increases. (b) Sensitive dependence on initial conditions. Top and bottom panels show two different views of a network started at very similar initial conditions. Top panel shows traces over time, with solid traces corresponding to original initial condition and dashed traces corresponding to same neurons in perturbed initial conditions. Bottom panel shows 3D state space plot of trajectories, with blue and red corresponding to original and perturbed initial conditions respectively, and network dynamics shown only for the first 20 units of time, with start and end points of the trajectories indicated. The perturbation to the initial condition of each neuron is uniformly drawn from $[0, 10^{-4}]$ and network gain, $g=10$. The original and slightly perturbed trajectories rapidly diverge and become uncorrelated, as can be seen from the separation between the solid and dashed traces at the top and between red and blue endpoints at the bottom. Network size $N=250$ in all plots. }
    \label{fig:chaotic_net}
\end{figure}

\section{Results}
\subsection{Chaotic network}
To model chaos in neural networks, we use a classic firing rate network model introduced by Sompolinsky, Crisanti \& Sommer \cite{sompolinsky1988}. The model network consists of $N$ randomly-connected neurons. The input (or synaptic current) to the $m$th neuron is $h_m(t)$, and the corresponding output (or activity or firing rate) is $\phi(h_m(t))$, where $\phi(\cdot)=\tanh(\cdot)$ is a sigmoidal activation function. The synaptic currents evolve over time as 
\begin{equation}\label{eq:somp_chaos_net}
\tau\frac{dh_m}{dt} = -h_m + g\sum_{n=1}^N W_{mn}\phi(h_n)
\end{equation}
where $\tau$ is the single neuron time constant and $W_{mn}$ is the recurrent connection strength from neuron $n$ to neuron $m$. The entries $W_{mn}$ are drawn IID from a Gaussian distribution with mean $0$ and variance $1/N$, where $g>0$ is an overall scaling parameter called the gain, and the factor of $1/N$ keeps the variance of the total drive constant as $N$ changes. If the activation function is applied elementwise, then this equation can be written as $\tau\frac{d\vc{h}}{dt} = -\vc{h} + gW\phi(\vc{h})$.

The gain parameter $g$ captures the overall strength of recurrent connections and determines the nature of the dynamics. For large $N$, the network shows a transition to chaos when $g=1$. When $g<1$, the network shows a single stable fixed point at $\vc{h} = \vc{0}$ and activity decays to this fixed point. When $g>1$, this fixed point is unstable and the network exhibits chaotic dynamics \cite{sompolinsky1988, kadmon2015transition}. 

Chaos leads to the exponential divergence of nearby trajectories and mixing, reflected in positive Lyapunov exponents and a positive entropy rate for $g>1$. Moreover the strength of chaos, as measured by the magnitude of the positive Lyapunov exponents and the entropy rate, grows with $g$ \cite{sompolinsky1988, engelken2023lyapunov}. Recent calculations of the Lyapunov spectrum and the entropy rate can be found in Engelken et al. \cite{engelken2023lyapunov}. Fig. \ref{fig:chaotic_net} shows time series and phase space plots of network activity for different values of $g$, highlighting the increasing strength of chaos as $g$ increases (panel a) and the sensitive dependence on initial conditions (panel b). 

This network is deterministic but, in the chaotic regime, it behaves pseudorandomly. In particular, pairwise correlations between neurons are small, network dynamics show a positive entropy rate, and information about the network's initial conditions becomes rapidly inaccessible to any finite level of resolution \cite{sompolinsky1988, clark2023dimension, engelken2023lyapunov}. Moreover, recent results show that entropy and the dimension of the chaotic attractor grow with network size ($N$), thus providing significant variability if $N$ is not too small \cite{clark2023dimension, engelken2023lyapunov} . 

In the rest of this study, we show that this pseudorandom activity allows the network to serve as an excellent source of variability for sampling from probabilistic models. We consider two architectures, one in which variability generated by the chaotic network is reshaped by a downstream network (Sections 3.2 and 3.3), and another in which random and structured connections coexist in the same recurrent network (Section 3.4). 

\subsection{Reshaping variability for generative modeling}
\label{sec:chaos_gan}
As a first example of how chaotic dynamics could be used to sample from generative models, we adapt a common generative modeling architecture in machine learning. A number of these architectures construct latent variable models for observed data $x$ of the form $P(x,z) = P(x|z)P(z)$, where $z$ is the latent variable, usually with a Gaussian prior. Or, put another way, these architectures learn to reshape simple (Gaussian-distributed) noise to match a desired data distribution \cite{Kingma2014, goodfellow2014generative, Rezende14, rezende2015variational, sohl2015deep, song2019generative, ho2020denoising}. Samples from the distribution can then be generated by drawing values of the noise and passing them through the network to turn them into samples from the data distribution.

These architectures can be adapted to instead reshape variability generated by a chaotic recurrent network. We demonstrate this idea by adapting a canonical generative modeling architecture, namely a Generative Adversarial Network (GAN) \cite{goodfellow2014generative}.



\begin{figure}[th]
    \begin{center}
            \includegraphics[width=\linewidth]{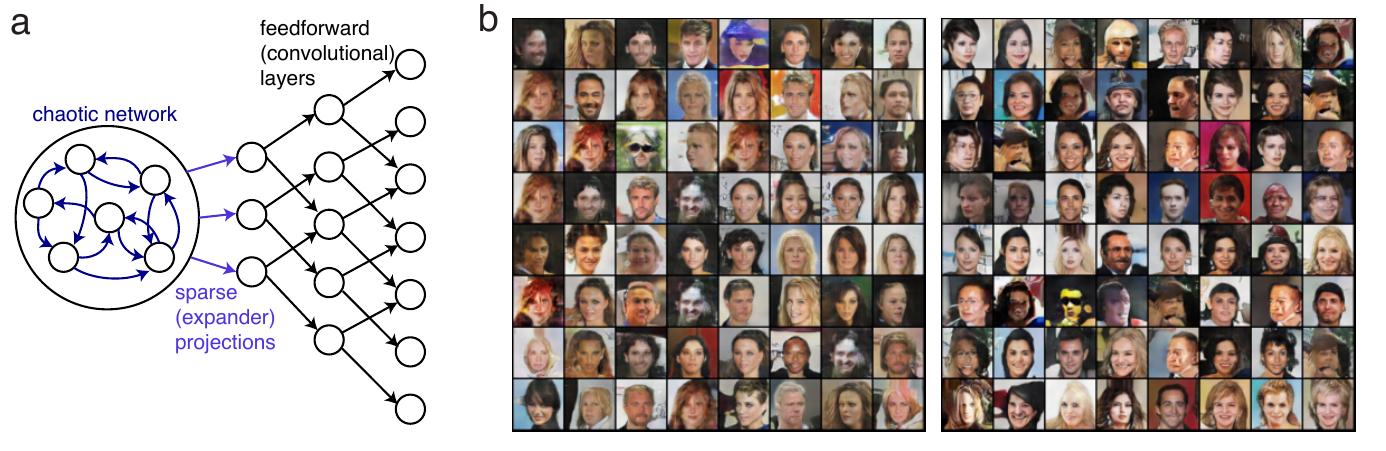}
    \end{center}
    \caption{{\bf Using chaos to sample from a Generative Adversarial Network.}
(a) Schematic of the generator. The recurrent network shown in the big circle has randomly-chosen recurrent connections (dark blue) and is in the chaotic regime. Projections out from the recurrent network are sparse and random, forming an expander graph (light blue). This activity is then propagated through a multilayer convolutional network to generate images. Blue connections are random and fixed during training, while black connections are trained. The architecture also includes a discriminator, which is not shown for simplicity. (b) Samples from the GAN trained on the CelebA dataset \cite{liu2015faceattributes}.}
    \label{fig:chaos_gen}
\end{figure}
 
A GAN consists of two networks: a generator network $G(\vc{z}; \vc{\theta_g})$ and a discriminator network $D(\vc{x}; \vc{\theta_d})$, with network parameters $\vc{\theta_g}$ and $\vc{\theta_d}$ respectively. The generator network $G(\vc{z}; \vc{\theta_g})$ maps some latent variables $\vc{z}\in \mathbb{R}^K$ into data space, where $K$ is the dimension of the latent space. The discriminator network $D(\vc{x}; \vc{\theta_d})$ takes as input $\vc{x}$ either from the dataset or created by the generator and outputs the probability that the input was drawn from the data (i.e., is real) as opposed to being the output of the generator (i.e., is generated). The two networks are engaged in a two-player game, in which the generator network attempts to generate outputs that resemble samples from the data distribution, while the discriminator network attempts to discriminate the generator output from the true data distribution. The cost function for the generator network thus measures how often it can fool the discriminator, and the cost for the discriminator network is the degree to which it can correctly identify real vs. generated data. 

In a traditional GAN, the inputs $\vc{z}$ to the generator are drawn from a prior noise distribution, typically a multivariate Gaussian distribution. Thus, the generator network generates samples by reshaping multivariate Gaussian distributed noise via a feedforward neural network. We modify this architecture and replace the source of noise by an $N$-neuron chaotic network from the previous section. Thus, we generate $\vc{z}$ using the process,
\begin{align}
\frac{d\vc{h}}{dt} &= -\vc{h} + gW^{\text{rec}}\phi(\vc{h}) \nonumber \\
\vc{z} &= W^{\text{out}}\phi(\vc{h}),
\end{align}
where the matrix $W^\text{out}$ is a sparse $K \times N$ readout matrix with binary entries. The weights within the chaotic network ($W^{\text{rec}}$) are initialized randomly using a zero-mean normal distribution with variance $g^2/N$. The readout matrix $W^\text{out}$ is chosen to be the adjacency matrix of a sparse bipartite expander graph. Such connectivity structures promote mixing and can be generated randomly. The results are not greatly sensitive to the architecture of the rest of the GAN (parameters for the remainder of the specific architecture we show in Figure \ref{fig:chaos_gen} are described in the Appendix, and are based on prior work \cite{radford2015unsupervised}). We freeze the weights $W^{\text{rec}}$ and $W^{\text{out}}$ and train the remaining parameters of the generator and discriminator using the standard GAN cost function.

Over training, the generator network learns to generate samples from the desired distribution, with variability internally generated by the chaotic network. Examples from a network trained on the CelebA dataset \cite{liu2015faceattributes} are shown in Fig. \ref{fig:chaos_gen}.

\subsection{Changing the sampling rate by modulating gain}
\begin{figure}[ht]
    \begin{center}
            \includegraphics[width=0.8\linewidth]{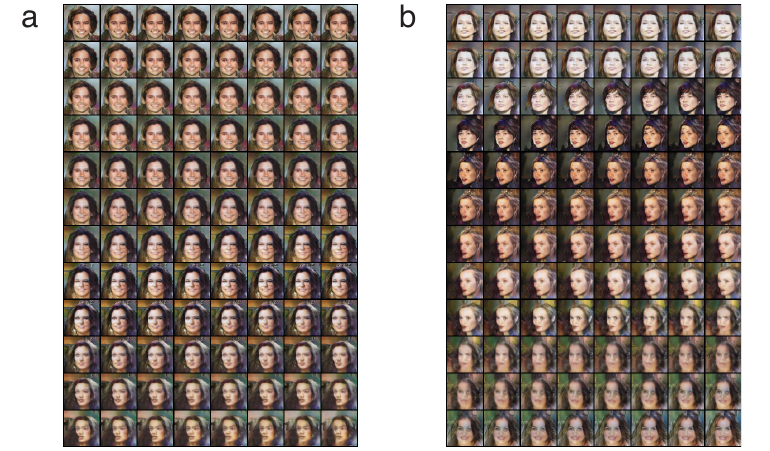}
    \end{center}
    \caption{{\bf Controlling the speed of sampling by modulating the gain.} 
(a) Consecutive samples from the network in Fig. \ref{fig:chaos_gen} at intervals of $0.15 \tau$ with gain set to $2$. (b) Consecutive samples from the network in Fig. \ref{fig:chaos_gen} at intervals of $0.15 \tau$ with gain set to $8$. Note that the network was trained with gain set to $4$ and not retrained to generate these samples. }
    \label{fig:speed_gain_mod}
\end{figure}
A neural sampling mechanism should allow some control over the speed at which new uncorrelated samples are presented. For example, a rat using sampling to explore different behavioral strategies should hold each sample for long enough to drive the appropriate behavior, potentially on the timescale of seconds to minutes. By contrast, a visual system that is using sampling to carry out Bayesian posterior inference in the service of facial recognition needs to be able to quickly draw a large number of samples to approximate the posterior distribution and recognize a face in under a second. Even within a single task context, the speed of sampling might vary depending on behavioral demands, speed-accuracy tradeoffs, and so on. Mechanisms that use intrinsic neuronal stochasticity to perform sampling typically lack a mechanism to control this sampling speed, since the statistics of the noise are determined by intrinsic biophysical timescales.

In a chaotic network like the one described by Eq. \ref{eq:somp_chaos_net} , however, the degree of chaos is controlled by the single scalar gain parameter $g$, which scales the strength of all recurrent connections. As $g$ increases, the dynamics more rapidly explores the space \cite{sompolinsky1988, engelken2023lyapunov}. By contrast, for small $g$ the network is barely chaotic. Thus, the value of $g$ provides a possible mechanism to control the sampling rate. 

To probe the effect of changing gain on the output samples, we consider the trained network shown in Fig. \ref{fig:chaos_gen}, which was trained for a gain of $4$ and switch the gain to either a lower or higher value without retraining the network. In both cases, we find that the network generalizes across gains and is able to generate good samples. Moreover, we find that sampling in the low-gain regime is slower than sampling in the high-gain regime, as shown in Fig. \ref{fig:speed_gain_mod}. Thus switching $g$ from a low to a high value acts to speed up sampling.  

The parameter $g$ has a number of natural biological interpretations, including as the global excitability or gain of single neurons. Such excitability is known to be state dependent, and there are a number of ways to control such excitability in a biologically-plausible way, including by neuromodulation. The gain of neural responses can also be controlled by baseline inputs that simply shift neurons' resting points along a nonlinear $f-I$ curve or by top-down inputs to distal dendrites. In sum, changing $g$ provides an easy and biologically-plausible knob with which to switch the speed of sampling from slow to fast.

\subsection{Combining structured dynamics with chaotic variability} 
The results presented so far separate the generation of variability by the chaotic network from a separate network that reshapes this variability. This separation has conceptual and computational advantages, and is common to a number of generative modeling architectures. Adopting this separation allows us to adapt these generative modeling architectures to harness chaotic dynamics. However, it is unclear if such a clean separation between the source and the use of variability could exist in the brain. In this section, we thus consider architectures in which structured dynamics and chaotic variability coexist in the same network.

\begin{figure}
    \begin{center}
            \includegraphics[width=\linewidth]{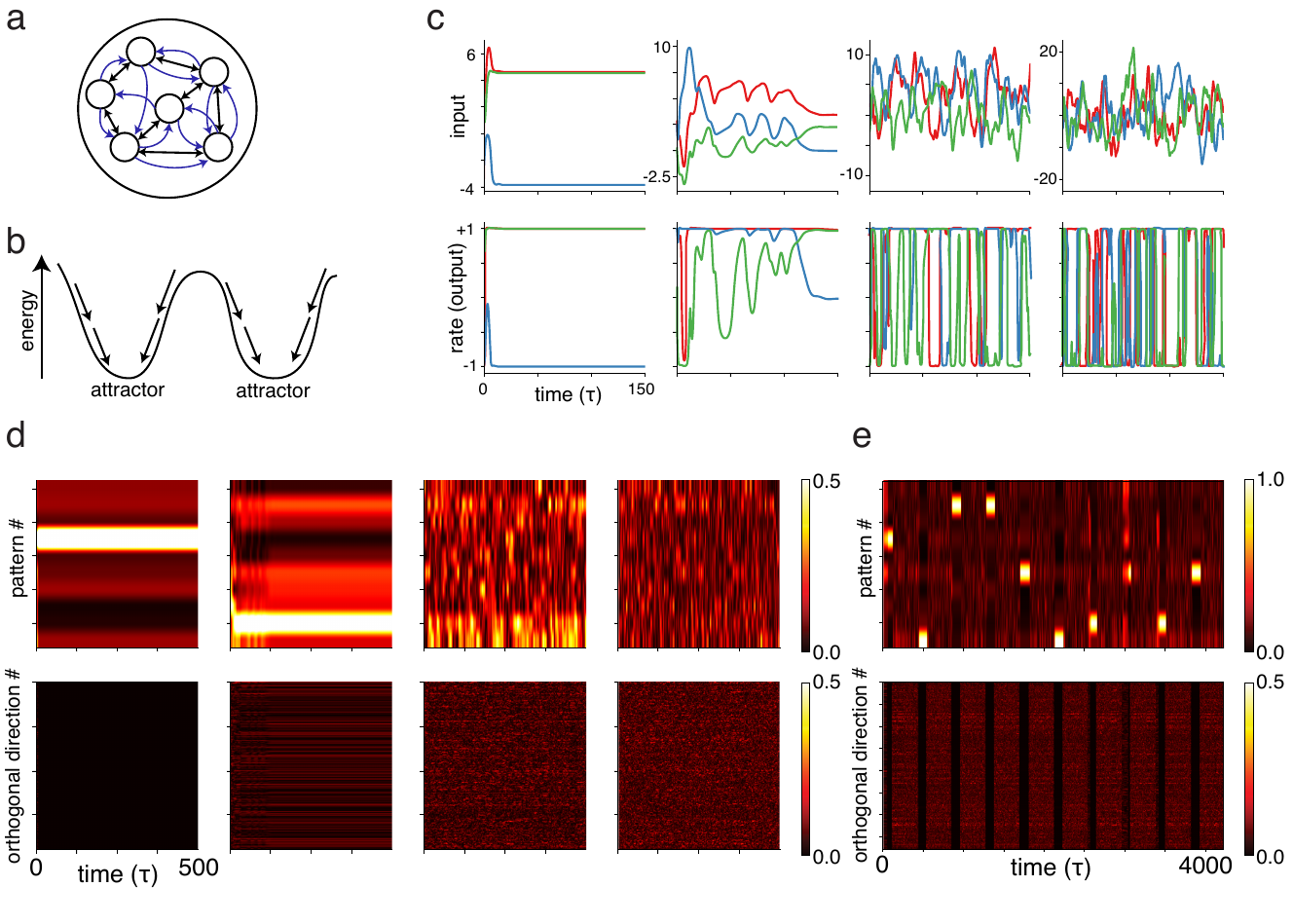}
    \end{center}
    \caption{{\bf Recurrent networks with combined structured and random connectivity.}
(a) Schematic of network with a combination of symmetric low-rank connectivity (black) and high-dimensional random connectivity (blue). (b) Schematic of energy-based dynamics in network when only symmetric low-rank component is present (i.e., gain of random connectivity is set to $0$). Dynamics moves the network state downhill in energy until it converges to an attractor state. Panels c, d apply to networks whose connectivity is given by $W = W_{\text{struct}} + gW_\text{rand}$, for some fixed $g$. Columns left to right show $g=0, 3, 5,$ and $8$. Note that the plot for each gain uses a different network initial condition and thus is not expected to converge to the same attractor state. (c) Top and bottom panels respectively show time-varying network input to and output from 3 neurons (i.e., as in Fig. 1a except network combines random and structured connectivity). Note the convergence to the stable attractor state in the first panel (when $g=0$). (d) Top panels show projection (normalized dot product) of network activity vector onto the $10$ patterns stored in the structured connectivity (i.e., the minima of the corresponding energy function), with each row showing a different projection. Bottom panels show projection of network activity onto $190$ orthogonal directions. For low gain, the activity is confined to the subspace spanned by stored patterns. As gain increases (i.e., left to right), network activity spends more time exploring orthogonal directions. (e) Results from network whose connectivity is given by $W = W_{\text{struct}} + g(t)W_\text{rand}$, where $g(t)$ cycles between a low value ($g=1$) and a high value ($g=8$). As in panel (d), top and bottom panels show projection onto stored patterns and orthogonal directions respectively. When $g$ is low, the network settles into one of the energy minima (seen as bright bands in top panel and dark bands in bottom panel). When $g$ is high, the network explores state space. Cycling $g$ allows the network to explore multiple energy minima. Note that the scale in the top panel is different, to highlight the sharp convergence to the energy minimum (i.e., overlap is close to perfect). }
    \label{fig:struct_rand}
\end{figure}

The framework suggested here is quite general. To make it concrete, we consider a simple $N$-neuron rate-based recurrent neural network model of the form described previously:
\begin{equation}
\tau\frac{d\vc{h}}{dt} = -\vc{h} + W\phi(\vc{h}).
\end{equation}
Unlike in the previous setting, we now assume that the connectivity matrix $W$ has both a structured component, $W_{\text{struct}}$, and a random component, $W_\text{rand}$, schematic shown in Fig. \ref{fig:struct_rand}a. We set $W = W_{\text{struct}} + g(t)W_\text{rand}$, where $g(t)$ is a possibly time-varying gain (see justification below for the inclusion of time dependence in $g$).

The structured component $W_{\text{struct}}$ encodes a set of patterns that the network has stored or preferred states that the network should sample from. In the absence of the random component (i.e., $g(t)=0$), the network activity is entirely determined by $W_{\text{struct}}$. To provide a simple and concrete example, we assume that $W_{\text{struct}}$ is set up so that in the absence of $W_\text{rand}$ the network is a continuous (classic) Hopfield network, a canonical energy-based architecture that has been extensively used to model memory \cite{hopfield1982neural}. In this setting, $W_{\text{struct}}$ is symmetric and defines an energy function. The network dynamics act to minimize this energy function---they move the network state downhill in energy until the network converges to one of a set of point attractor states, determined by $W_{\text{struct}}$ (shown in the schematic in Fig. \ref{fig:struct_rand}b).  Given a set of $K$ desired attractor states, $\{\vc{\xi_1}, \vc{\xi_2}, \hdots, \vc{\xi_K}\}$, we construct $W_{\text{struct}}$ using the so-called outer product learning rule \cite{hopfield1982neural}. That is, $W_{\text{struct}}= \sum_{k=1}^K \vc{\xi_k}\vc{\xi_k}^T$ and is rank $K$. As long as $K$ is not too large and the patterns are not highly correlated, this construction guarantees that the $K$ patterns $\vc{\xi_k}$ are close to minima (stable attractors) of the energy function. Note that the patterns can be made exact minima using the slightly more complicated pseudoinverse rule \cite{kanter1987associative}, but for our minimal example we choose the simpler setting. 

The random component $W_\text{rand}$ is identical to the matrix from Eq. \ref{eq:somp_chaos_net}, and has IID Gaussian entries with mean $0$ and variance $1/N$, as before. In the absence of the structured component and for $g>1$ (assuming $N$ sufficiently large), the network dynamics are thus chaotic, converging to and then rapidly exploring the chaotic attractor, and showing pseudorandom, highly variable activity determined by the invariant measure on the chaotic attractor. 

Next consider the combined system, with connectivity matrix $W = W_{\text{struct}} + g(t)W_\text{rand}$ consisting of a structured and a random part. The contribution of the random part is scaled by a gain, $g(t)$, that controls the balance of structure and randomness. There are two natural settings in which this system can be used to sample from a probability distribution shaped by $W_{\text{struct}}$.

First, if $g$ is fixed at some particular value then the resulting dynamics interpolate between the gradient-based (energy-minimization) dynamics of $W_{\text{struct}}$, and the chaotic, highly mixing dynamics of $W_\text{rand}$. In Fig. \ref{fig:struct_rand}c we show neural activity from four different realizations of the gain, highlighting this increasing variability. The $g=0$ case corresponds to pure gradient-based dynamics, with convergence to a stable attractor state, visible in the leftmost panel of Fig. \ref{fig:struct_rand}c. As $g$ increases (left to right in Fig. \ref{fig:struct_rand}c), the random component increasingly outweighs the structured component and the dynamics approach the chaotic network of Eq. \ref{eq:somp_chaos_net}. 

To characterize the overlap of population activity with the $K$ patterns encoded in the structured connectivity, we compute the normalized dot product between the population activity vector and each of these patterns, shown in the top panel of Fig. \ref{fig:struct_rand}d. To contrast this overlap with the remaining $N-K$ directions in state space, we construct an orthogonal basis for the $N-K$ dimensional non-pattern subspace and compute the normalized dot product of the population vector with each of these basis vectors, shown in the bottom panel of Fig. \ref{fig:struct_rand}d. As can be seen moving left to right, for low gain the population activity overlaps with the stored patterns and the degree of overlap decreases as gain increases. The correlation between the gain and the overlap with the stored patterns is very strong and highly significant ($r=-0.95$, $p<10^{-16}$).

Alternatively, instead of being fixed, the gain $g$ could vary over time. As described in the previous section, there are a number of biologically-plausible ways to modulate gain or excitability, including neuromodulation and dendritic input. If the random connections are of a different type or in a different layer from the structured ones, then it is possible that the overall strength of these connections could be modulated differently from the structured connections. In this case, $g(t)$ would be a time-varying quantity that dynamically controls the relative impact of the structured and random connections. 

The ability to change $g(t)$ could thus switch the network between a regime with attractor-like dynamics, where it draws samples from the vicinity of a stored pattern (low $g$), and a regime that allows the network to explore the space and move to a new attractor basin (high $g$). In Fig. \ref{fig:struct_rand}e we show results from such a network with gain alternating between a low and a high value. The network correspondingly shows alternation of attractor-like and mixing-driven dynamical regimes as $g$ varies. This alternation is visible in the plots of overlap between the network state and either stored patterns or orthogonal directions. When gain is low, the network settles into one of the attractor states (strong overlap with one of the stored patterns, visible as bright bands in top panel of Fig. \ref{fig:struct_rand}e). When gain is high, the network explores the space (low overlap with the stored patterns), and then settles into another attractor state when gain is reduced again (compare consecutive bright bands in top panel of Fig. \ref{fig:struct_rand}e). Thus, such a network effectively samples from the low-energy states of the energy function given by the structured connectivity, with the random connectivity driving transitions between low-energy states.  

\section{Discussion}
In this study we show how chaotic neural network dynamics can be used to sample from probabilistic models using two network architectures. First, we show how variability from a chaotic network can be reshaped to match a desired distribution in the context of a classic generative modeling architecture. Second, we show how chaos from random connectivity can be used to sample from a set of preferred states in a recurrent network with a combination of structured and random connectivity. We also highlight how gain modulation can be used to control both the rate of sampling and the balance between structured and chaotic dynamics. 

Most prior work has focused on using single neuron or synapse-level sources of stochasticity to drive sampling. However, it is unclear whether this stochasticity is sufficient to drive good sampling in the absence of network-level mechanisms to amplify small differences. First, in the absence of network effects single neurons can  be quite reliable and thus single neuron sources of stochasticity could be small \cite{mainen1995reliability}. Second, given that neocortical neurons receive a large number of inputs, unless single neuron stochasticity is strongly correlated it is likely to average out at the population level. By contrast, chaotic network dynamics has a number of appealing properties as a source of variability for sampling. In particular, the variability in these networks (as measured by entropy) does not average out and instead grows with network size \cite{sompolinsky1988, clark2023dimension, engelken2023lyapunov}. Moreover, chaotic dynamics contains exponentially unstable directions, thus allowing trajectories to explore a space rapidly when compared to noisy but stable dynamics. Network-level mechanisms also allow the potential to control the speed of sampling, possibly by modulating overall excitability or background input. More generally, chaotic mechanisms are likely to coexist with and potentially amplify neuron- and synapse-intrinsic sources of stochasticity. 
 
We note that using a chaotic neural network to generate variability is not an efficient strategy on a standard serial computer architecture, especially given access to fast pseudorandom number generators, and thus is a limitation of our architectures for traditional machine learning applications. However, such a strategy might be appealing in a biological system, which is naturally parallel and highly variable. Such a strategy might also be appealing for neuromorphic computing applications, in which devices are once again naturally parallel and components often show significant intrinsic variability that could be reshaped for sampling. 

The framework we present in the first half of this study separates the generation and reshaping of variability, much like common generative modeling architectures in machine learning \cite{Kingma2014, goodfellow2014generative, Rezende14, rezende2015variational, sohl2015deep, song2019generative, ho2020denoising}. It also resembles reservoir computing architectures, in that complex high-dimensional dynamics are provided by one network as input to another \cite{maass2002real}. In our case, the reservoir network acts much like a pseudorandom number generator and could be optimized to generate samples with a desired sampling speed and degree of correlation. This separation of networks is computationally efficient and conceptually clean, and might again be appealing for neuromorphic computing applications. But the separation is also a limitation of the first architecture in that it is likely biologically unrealistic. 


The second architecture presented in this study considers networks in which the probability distribution is directly encoded in the weights of the recurrent connections by a combination of structured and random connectivity. As opposed to the more complex generative model presented in the first architecture, these networks sample from quite simple probabilistic models, consisting of a set of preferred network activity patterns. Thus, these networks are currently limited in their capabilities. However, we envisage them as proof of principle for the more general idea of combining gain-modulated random connectivity with structured connectivity, and thus as the first step in constructing more complex models using this architecture. 

Networks with a combination of structured low-rank and random connectivity are an active area of current research, with increasing support both for the presence of such network structures in the brain and for their computational utility \cite{landau2018coherent, mastrogiuseppe2018linking, schuessler2020dynamics}. Notably, Landau \& Sompolinsky \cite{landau2018coherent} show how the addition of low-rank connectivity to a random network can shape the resulting chaotic variability (though they do not consider the problem of sampling). By contrast, a second line of work studies how low-dimensional structure can emerge from low-rank connectivity, even when embedded in background high-dimensional random connectivity \cite{mastrogiuseppe2018linking, schuessler2020dynamics}. To our knowledge, none of these prior studies explicitly considers these networks as probabilistic samplers. Prior work on such networks moreover typically considers a regime where either the structured or random component dominates, and have not considered dynamically changing the balance of structured and random connectivity. Our results suggest that the ability to dynamically change this balance is a powerful computational mechanism, allowing the same network to potentially do both stereotyped and probabilistic computation and to adjust the degree of exploration based on task demands. It is an open empirical question whether connections are indeed dynamically modulated this way and if this is a possible mechanism to modulate transitions between stability and flexible exploration. Thus this speculative mechanism is both a potential limitation of our study and an exciting possibility to explore. 

\begin{ack}
This material is based upon work supported by the Air Force Office of Scientific Research (AFOSR) under award number FA9550-22-1-0532. The authors were benefited by participating in the activities of the UC Davis TETRAPODS Institute of Data Science, which has been funded by the NSF TRIPODS grant CCF-1934568. 
%
%
\end{ack}
\clearpage
\bibliography{chaos_sampling}

\begin{thebibliography}{10}
\expandafter\ifx\csname url\endcsname\relax
  \def\url#1{\texttt{#1}}\fi
\expandafter\ifx\csname urlprefix\endcsname\relax\def\urlprefix{URL }\fi
\providecommand{\bibinfo}[2]{#2}
\providecommand{\eprint}[2][]{\url{#2}}

\bibitem{sompolinsky1988}
\bibinfo{author}{Sompolinsky, H.}, \bibinfo{author}{Crisanti, A.} \&
  \bibinfo{author}{Sommers, H.-J.}
\newblock \bibinfo{title}{Chaos in random neural networks}.
\newblock \emph{\bibinfo{journal}{Physical review letters}}
  \textbf{\bibinfo{volume}{61}}, \bibinfo{pages}{259} (\bibinfo{year}{1988}).

\bibitem{kadmon2015transition}
\bibinfo{author}{Kadmon, J.} \& \bibinfo{author}{Sompolinsky, H.}
\newblock \bibinfo{title}{Transition to chaos in random neuronal networks}.
\newblock \emph{\bibinfo{journal}{Physical Review X}}
  \textbf{\bibinfo{volume}{5}}, \bibinfo{pages}{041030} (\bibinfo{year}{2015}).

\bibitem{aljadeff2015transition}
\bibinfo{author}{Aljadeff, J.}, \bibinfo{author}{Stern, M.} \&
  \bibinfo{author}{Sharpee, T.}
\newblock \bibinfo{title}{Transition to chaos in random networks with
  cell-type-specific connectivity}.
\newblock \emph{\bibinfo{journal}{Physical review letters}}
  \textbf{\bibinfo{volume}{114}}, \bibinfo{pages}{088101}
  (\bibinfo{year}{2015}).

\bibitem{harish2015asynchronous}
\bibinfo{author}{Harish, O.} \& \bibinfo{author}{Hansel, D.}
\newblock \bibinfo{title}{Asynchronous rate chaos in spiking neuronal
  circuits}.
\newblock \emph{\bibinfo{journal}{PLoS computational biology}}
  \textbf{\bibinfo{volume}{11}}, \bibinfo{pages}{e1004266}
  (\bibinfo{year}{2015}).

\bibitem{london2010sensitivity}
\bibinfo{author}{London, M.}, \bibinfo{author}{Roth, A.},
  \bibinfo{author}{Beeren, L.}, \bibinfo{author}{H{\"a}usser, M.} \&
  \bibinfo{author}{Latham, P.~E.}
\newblock \bibinfo{title}{Sensitivity to perturbations in vivo implies high
  noise and suggests rate coding in cortex}.
\newblock \emph{\bibinfo{journal}{Nature}} \textbf{\bibinfo{volume}{466}},
  \bibinfo{pages}{123--127} (\bibinfo{year}{2010}).

\bibitem{crutchfield1990computation}
\bibinfo{author}{Crutchfield, J.~P.} \& \bibinfo{author}{Young, K.}
\newblock \bibinfo{title}{Computation at the onset of chaos}.
\newblock \emph{\bibinfo{journal}{Entropy, complexity, and the physics of
  information}} \textbf{\bibinfo{volume}{8}}, \bibinfo{pages}{223}
  (\bibinfo{year}{1990}).

\bibitem{bertschinger2004real}
\bibinfo{author}{Bertschinger, N.} \& \bibinfo{author}{Natschl{\"a}ger, T.}
\newblock \bibinfo{title}{Real-time computation at the edge of chaos in
  recurrent neural networks}.
\newblock \emph{\bibinfo{journal}{Neural computation}}
  \textbf{\bibinfo{volume}{16}}, \bibinfo{pages}{1413--1436}
  (\bibinfo{year}{2004}).

\bibitem{toyoizumi2011beyond}
\bibinfo{author}{Toyoizumi, T.} \& \bibinfo{author}{Abbott, L.~F.}
\newblock \bibinfo{title}{Beyond the edge of chaos: Amplification and temporal
  integration by recurrent networks in the chaotic regime}.
\newblock \emph{\bibinfo{journal}{Physical Review E}}
  \textbf{\bibinfo{volume}{84}}, \bibinfo{pages}{051908}
  (\bibinfo{year}{2011}).

\bibitem{sussillo2009generating}
\bibinfo{author}{Sussillo, D.} \& \bibinfo{author}{Abbott, L.~F.}
\newblock \bibinfo{title}{Generating coherent patterns of activity from chaotic
  neural networks}.
\newblock \emph{\bibinfo{journal}{Neuron}} \textbf{\bibinfo{volume}{63}},
  \bibinfo{pages}{544--557} (\bibinfo{year}{2009}).

\bibitem{laje2013robust}
\bibinfo{author}{Laje, R.} \& \bibinfo{author}{Buonomano, D.~V.}
\newblock \bibinfo{title}{Robust timing and motor patterns by taming chaos in
  recurrent neural networks}.
\newblock \emph{\bibinfo{journal}{Nature neuroscience}}
  \textbf{\bibinfo{volume}{16}}, \bibinfo{pages}{925--933}
  (\bibinfo{year}{2013}).

\bibitem{terada2024chaotic}
\bibinfo{author}{Terada, Y.} \& \bibinfo{author}{Toyoizumi, T.}
\newblock \bibinfo{title}{Chaotic neural dynamics facilitate probabilistic
  computations through sampling}.
\newblock \emph{\bibinfo{journal}{Proceedings of the National Academy of
  Sciences}} \textbf{\bibinfo{volume}{121}}, \bibinfo{pages}{e2312992121}
  (\bibinfo{year}{2024}).

\bibitem{ackley1985learning}
\bibinfo{author}{Ackley, D.~H.}, \bibinfo{author}{Hinton, G.~E.} \&
  \bibinfo{author}{Sejnowski, T.~J.}
\newblock \bibinfo{title}{A learning algorithm for boltzmann machines}.
\newblock \emph{\bibinfo{journal}{Cognitive science}}
  \textbf{\bibinfo{volume}{9}}, \bibinfo{pages}{147--169}
  (\bibinfo{year}{1985}).

\bibitem{dayan1995helmholtz}
\bibinfo{author}{Dayan, P.}, \bibinfo{author}{Hinton, G.~E.},
  \bibinfo{author}{Neal, R.~M.} \& \bibinfo{author}{Zemel, R.~S.}
\newblock \bibinfo{title}{The helmholtz machine}.
\newblock \emph{\bibinfo{journal}{Neural computation}}
  \textbf{\bibinfo{volume}{7}}, \bibinfo{pages}{889--904}
  (\bibinfo{year}{1995}).

\bibitem{buesing2011neural}
\bibinfo{author}{Buesing, L.}, \bibinfo{author}{Bill, J.},
  \bibinfo{author}{Nessler, B.} \& \bibinfo{author}{Maass, W.}
\newblock \bibinfo{title}{Neural dynamics as sampling: a model for stochastic
  computation in recurrent networks of spiking neurons}.
\newblock \emph{\bibinfo{journal}{PLoS computational biology}}
  \textbf{\bibinfo{volume}{7}}, \bibinfo{pages}{e1002211}
  (\bibinfo{year}{2011}).

\bibitem{hennequin2014fast}
\bibinfo{author}{Hennequin, G.}, \bibinfo{author}{Aitchison, L.} \&
  \bibinfo{author}{Lengyel, M.}
\newblock \bibinfo{title}{Fast sampling-based inference in balanced neuronal
  networks}.
\newblock \emph{\bibinfo{journal}{Advances in neural information processing
  systems}} \textbf{\bibinfo{volume}{27}} (\bibinfo{year}{2014}).

\bibitem{savin2014spatio}
\bibinfo{author}{Savin, C.} \& \bibinfo{author}{Den{\`e}ve, S.}
\newblock \bibinfo{title}{Spatio-temporal representations of uncertainty in
  spiking neural networks}.
\newblock \emph{\bibinfo{journal}{Advances in neural information processing
  systems}} \textbf{\bibinfo{volume}{27}} (\bibinfo{year}{2014}).

\bibitem{orban2016neural}
\bibinfo{author}{Orb{\'a}n, G.}, \bibinfo{author}{Berkes, P.},
  \bibinfo{author}{Fiser, J.} \& \bibinfo{author}{Lengyel, M.}
\newblock \bibinfo{title}{Neural variability and sampling-based probabilistic
  representations in the visual cortex}.
\newblock \emph{\bibinfo{journal}{Neuron}} \textbf{\bibinfo{volume}{92}},
  \bibinfo{pages}{530--543} (\bibinfo{year}{2016}).

\bibitem{gershman2017complex}
\bibinfo{author}{Gershman, S.~J.} \& \bibinfo{author}{Beck, J.~M.}
\newblock \bibinfo{title}{Complex probabilistic inference: From cognition to
  neural computation}.
\newblock \emph{\bibinfo{journal}{Computational models of brain and behavior}}
  \bibinfo{pages}{453--466} (\bibinfo{year}{2017}).

\bibitem{echeveste2020cortical}
\bibinfo{author}{Echeveste, R.}, \bibinfo{author}{Aitchison, L.},
  \bibinfo{author}{Hennequin, G.} \& \bibinfo{author}{Lengyel, M.}
\newblock \bibinfo{title}{Cortical-like dynamics in recurrent circuits
  optimized for sampling-based probabilistic inference}.
\newblock \emph{\bibinfo{journal}{Nature neuroscience}}
  \textbf{\bibinfo{volume}{23}}, \bibinfo{pages}{1138--1149}
  (\bibinfo{year}{2020}).

\bibitem{Kingma2014}
\bibinfo{author}{Kingma, D.~P.} \& \bibinfo{author}{Welling, M.}
\newblock \bibinfo{title}{Auto-encoding variational bayes}.
\newblock \emph{\bibinfo{journal}{Proceedings of the International Conference
  on Learning Representations (ICLR)}}  (\bibinfo{year}{2014}).

\bibitem{goodfellow2014generative}
\bibinfo{author}{Goodfellow, I.} \emph{et~al.}
\newblock \bibinfo{title}{Generative adversarial nets}.
\newblock \emph{\bibinfo{journal}{Advances in neural information processing
  systems}} \textbf{\bibinfo{volume}{27}} (\bibinfo{year}{2014}).

\bibitem{Rezende14}
\bibinfo{author}{Rezende, D.~J.}, \bibinfo{author}{Mohamed, S.} \&
  \bibinfo{author}{Wierstra, D.}
\newblock \bibinfo{title}{Stochastic backpropagation and approximate inference
  in deep generative models}.
\newblock \emph{\bibinfo{journal}{Proceedings of the 31st International
  Conference on Machine Learning, pp. 1278–1286}}  (\bibinfo{year}{2014}).

\bibitem{rezende2015variational}
\bibinfo{author}{Rezende, D.} \& \bibinfo{author}{Mohamed, S.}
\newblock \bibinfo{title}{Variational inference with normalizing flows}.
\newblock In \emph{\bibinfo{booktitle}{International conference on machine
  learning}}, \bibinfo{pages}{1530--1538} (\bibinfo{organization}{PMLR},
  \bibinfo{year}{2015}).

\bibitem{sohl2015deep}
\bibinfo{author}{Sohl-Dickstein, J.}, \bibinfo{author}{Weiss, E.},
  \bibinfo{author}{Maheswaranathan, N.} \& \bibinfo{author}{Ganguli, S.}
\newblock \bibinfo{title}{Deep unsupervised learning using nonequilibrium
  thermodynamics}.
\newblock In \emph{\bibinfo{booktitle}{International conference on machine
  learning}}, \bibinfo{pages}{2256--2265} (\bibinfo{organization}{PMLR},
  \bibinfo{year}{2015}).

\bibitem{song2019generative}
\bibinfo{author}{Song, Y.} \& \bibinfo{author}{Ermon, S.}
\newblock \bibinfo{title}{Generative modeling by estimating gradients of the
  data distribution}.
\newblock \emph{\bibinfo{journal}{Advances in neural information processing
  systems}} \textbf{\bibinfo{volume}{32}} (\bibinfo{year}{2019}).

\bibitem{ho2020denoising}
\bibinfo{author}{Ho, J.}, \bibinfo{author}{Jain, A.} \&
  \bibinfo{author}{Abbeel, P.}
\newblock \bibinfo{title}{Denoising diffusion probabilistic models}.
\newblock \emph{\bibinfo{journal}{Advances in neural information processing
  systems}} \textbf{\bibinfo{volume}{33}}, \bibinfo{pages}{6840--6851}
  (\bibinfo{year}{2020}).

\bibitem{clark2023dimension}
\bibinfo{author}{Clark, D.~G.}, \bibinfo{author}{Abbott, L.} \&
  \bibinfo{author}{Litwin-Kumar, A.}
\newblock \bibinfo{title}{Dimension of activity in random neural networks}.
\newblock \emph{\bibinfo{journal}{Physical Review Letters}}
  \textbf{\bibinfo{volume}{131}}, \bibinfo{pages}{118401}
  (\bibinfo{year}{2023}).

\bibitem{engelken2023lyapunov}
\bibinfo{author}{Engelken, R.}, \bibinfo{author}{Wolf, F.} \&
  \bibinfo{author}{Abbott, L.~F.}
\newblock \bibinfo{title}{Lyapunov spectra of chaotic recurrent neural
  networks}.
\newblock \emph{\bibinfo{journal}{Physical Review Research}}
  \textbf{\bibinfo{volume}{5}}, \bibinfo{pages}{043044} (\bibinfo{year}{2023}).

\bibitem{eckmann1985ergodic}
\bibinfo{author}{Eckmann, J.-P.} \& \bibinfo{author}{Ruelle, D.}
\newblock \bibinfo{title}{Ergodic theory of chaos and strange attractors}.
\newblock \emph{\bibinfo{journal}{Reviews of modern physics}}
  \textbf{\bibinfo{volume}{57}}, \bibinfo{pages}{617} (\bibinfo{year}{1985}).

\bibitem{gilpin2024generative}
\bibinfo{author}{Gilpin, W.}
\newblock \bibinfo{title}{Generative learning for nonlinear dynamics}.
\newblock \emph{\bibinfo{journal}{Nature Reviews Physics}}
  \bibinfo{pages}{1--13} (\bibinfo{year}{2024}).

\bibitem{naruse2019generative}
\bibinfo{author}{Naruse, M.} \emph{et~al.}
\newblock \bibinfo{title}{Generative adversarial network based on chaotic time
  series}.
\newblock \emph{\bibinfo{journal}{Scientific reports}}
  \textbf{\bibinfo{volume}{9}}, \bibinfo{pages}{12963} (\bibinfo{year}{2019}).

\bibitem{jordan2019deterministic}
\bibinfo{author}{Jordan, J.} \emph{et~al.}
\newblock \bibinfo{title}{Deterministic networks for probabilistic computing}.
\newblock \emph{\bibinfo{journal}{Scientific reports}}
  \textbf{\bibinfo{volume}{9}}, \bibinfo{pages}{18303} (\bibinfo{year}{2019}).

\bibitem{liu2015faceattributes}
\bibinfo{author}{Liu, Z.}, \bibinfo{author}{Luo, P.}, \bibinfo{author}{Wang,
  X.} \& \bibinfo{author}{Tang, X.}
\newblock \bibinfo{title}{Deep learning face attributes in the wild}.
\newblock In \emph{\bibinfo{booktitle}{Proceedings of International Conference
  on Computer Vision (ICCV)}} (\bibinfo{year}{2015}).

\bibitem{radford2015unsupervised}
\bibinfo{author}{Radford, A.}, \bibinfo{author}{Metz, L.} \&
  \bibinfo{author}{Chintala, S.}
\newblock \bibinfo{title}{Unsupervised representation learning with deep
  convolutional generative adversarial networks}.
\newblock \emph{\bibinfo{journal}{arXiv preprint arXiv:1511.06434}}
  (\bibinfo{year}{2015}).

\bibitem{hopfield1982neural}
\bibinfo{author}{Hopfield, J.~J.}
\newblock \bibinfo{title}{Neural networks and physical systems with emergent
  collective computational abilities.}
\newblock \emph{\bibinfo{journal}{Proceedings of the national academy of
  sciences}} \textbf{\bibinfo{volume}{79}}, \bibinfo{pages}{2554--2558}
  (\bibinfo{year}{1982}).

\bibitem{kanter1987associative}
\bibinfo{author}{Kanter, I.} \& \bibinfo{author}{Sompolinsky, H.}
\newblock \bibinfo{title}{Associative recall of memory without errors}.
\newblock \emph{\bibinfo{journal}{Physical Review A}}
  \textbf{\bibinfo{volume}{35}}, \bibinfo{pages}{380} (\bibinfo{year}{1987}).

\bibitem{mainen1995reliability}
\bibinfo{author}{Mainen, Z.~F.} \& \bibinfo{author}{Sejnowski, T.~J.}
\newblock \bibinfo{title}{Reliability of spike timing in neocortical neurons}.
\newblock \emph{\bibinfo{journal}{Science}} \textbf{\bibinfo{volume}{268}},
  \bibinfo{pages}{1503--1506} (\bibinfo{year}{1995}).

\bibitem{maass2002real}
\bibinfo{author}{Maass, W.}, \bibinfo{author}{Natschl{\"a}ger, T.} \&
  \bibinfo{author}{Markram, H.}
\newblock \bibinfo{title}{Real-time computing without stable states: A new
  framework for neural computation based on perturbations}.
\newblock \emph{\bibinfo{journal}{Neural computation}}
  \textbf{\bibinfo{volume}{14}}, \bibinfo{pages}{2531--2560}
  (\bibinfo{year}{2002}).

\bibitem{landau2018coherent}
\bibinfo{author}{Landau, I.~D.} \& \bibinfo{author}{Sompolinsky, H.}
\newblock \bibinfo{title}{Coherent chaos in a recurrent neural network with
  structured connectivity}.
\newblock \emph{\bibinfo{journal}{PLoS computational biology}}
  \textbf{\bibinfo{volume}{14}}, \bibinfo{pages}{e1006309}
  (\bibinfo{year}{2018}).

\bibitem{mastrogiuseppe2018linking}
\bibinfo{author}{Mastrogiuseppe, F.} \& \bibinfo{author}{Ostojic, S.}
\newblock \bibinfo{title}{Linking connectivity, dynamics, and computations in
  low-rank recurrent neural networks}.
\newblock \emph{\bibinfo{journal}{Neuron}} \textbf{\bibinfo{volume}{99}},
  \bibinfo{pages}{609--623} (\bibinfo{year}{2018}).

\bibitem{schuessler2020dynamics}
\bibinfo{author}{Schuessler, F.}, \bibinfo{author}{Dubreuil, A.},
  \bibinfo{author}{Mastrogiuseppe, F.}, \bibinfo{author}{Ostojic, S.} \&
  \bibinfo{author}{Barak, O.}
\newblock \bibinfo{title}{Dynamics of random recurrent networks with correlated
  low-rank structure}.
\newblock \emph{\bibinfo{journal}{Physical Review Research}}
  \textbf{\bibinfo{volume}{2}}, \bibinfo{pages}{013111} (\bibinfo{year}{2020}).

\end{thebibliography}


\appendix

\section{Appendix / supplemental material}

\subsection{Parameters for numerical experiments}
All simulations were carried out using Python and standard Python libraries (Pytorch, Numpy, Matplotlib) and run on a local cluster with A6000 GPUs. All networks could be trained to good performance in under 2 days on a single GPU.

\paragraph{Figure 1} The network consisted of $N=250$ neurons, with entries of the recurrent connectivity matrix drawn IID from a normal distribution with mean $0$ and variance $g^2/N$. In the plots $g$ took the values $2$, $4$, and $10$. For the plots of sensitive dependence on initial conditions, we considered the $g=10$ case. We took a randomly chosen initial condition and compared it to a perturbed version, where a uniform perturbation in the range $[0, 10^{-4}]$ was independently applied to the initial value for each neuron. 

\paragraph{Figure 2} As for Figure 1, the recurrent network that provided input to the GAN consisted of $N=250$ neurons, with entries of the recurrent connectivity matrix drawn IID from a normal distribution with mean $0$ and variance $g^2/N$. We set the gain $g=4$. A random subset of $100$ neurons was then used as the latent activation of the GAN generator. For the rest of the GAN, we used a standard architecture based on Radford et al. (2015) \cite{radford2015unsupervised}, which we describe for completeness. The generator takes as input the $100$-dimensional latent vector from the chaotic network and passes it through five convolutional transpose layers, each with batch normalization and ReLU activation before a final Tanh layer to produce $64 \times 64$ $3$ color channel images. The discriminator takes as input $64 \times 64$ $3$ color channel images, passes them through five convolutional layers, each with batch normalization and ReLU activation, before a final sigmoidal layer that outputs the probability that the input was from the real data distribution. The network was trained on the standard GAN loss function using the Adam optimizer. 
 
\paragraph{Figure 3} To generate the plots in Figure 3, we took the architecture trained for Figure 2 and rescaled the weights in the recurrent network to have the appropriate gain. For panel (a), we divided weights by $\sqrt{2}$ and for panel (b), we multiplied them by $\sqrt{2}$ (i.e., halve or double the variance). Other network parameters were not changed. We then ran the recurrent network over time and recorded samples at intervals of $0.15\tau$ from a single trajectory, to visualize how quickly the generated images changed over time.

\paragraph{Figure 4} For the simulations in Figure 4, we used networks of size $N=200$. We generated $10$ random patterns with entries in $\{-1, +1\}$ to store in the network. The structured component of the connectivity was the outer product of the $N \times 10$ matrix of patterns. For the random component of the connectivity, we first generated a $200 \times 200$ random matrix with zero mean, unit variance Gaussian entries and then rescaled it by the appropriate fixed or time-varying gain. In panels c, d, the gain took the values $0$, $3$, $5$, and $8$ (moving left to right). In panel e, the gain took the value of $1$ for a time uniformly distributed between $100\tau$ and $120\tau$ before switching to a gain of $8$ for a time uniformly distributed between $300\tau$ and $320\tau$. To measure the overlaps with the stored patterns in panels d and e, we computed the dot product of the population vector with each stored pattern divided by the lengths of each vector. To measure the overlap with the subspace orthogonal to that spanned by the patterns, we construct an orthogonal basis for that subspace using the SVD and project onto each of the basis vectors. To measure how the overlap with the pattern subspace and the orthogonal subspace changed as a function of $g$, we first computed the norm of the projection into each subspace at each moment in time and then computed the correlation of these norms with the corresponding value of gain. To evaluate significance, we used a $t$-test to compare the estimated slope to $0$ (implemented by SciPy's linregress function).


\end{document}